\documentclass[twocolumn,twoside,showpacs,preprintnumbers,amsmath,amssymb]{revtex4}

\usepackage{graphicx}
\usepackage{dcolumn}
\usepackage{bm}

\newcommand {\bc} {\begin{center}}
\newcommand {\ec} {\end{center}}
\newcommand {\be} {\begin{equation}}
\newcommand {\ee} {\end{equation}}

\begin{document}

\title{Vacancy-Impurity Nanoclusters in Solid Solutions $\mathrm{^3He -
^4He}$}

\author{Dimitar I. Pushkarov}
 \altaffiliation[
 On leave from ] {Institute of Solid State Physics
 Bulgarian Academy of Sciences,
 1784 Sofia, Bulgaria}
 \affiliation{Laboratoire de Physique Th\'eorique et Mod\'elisation, Universit\'e de
  Cergy-Pontoise, France}
 \email{E-mail: dipushk@issp.bas.bg}
\date{April 29, 2004 
}

\begin{abstract}
 The existence of vacancy--impurity clusters due to quantum properties
of vacancies in phase separated solid solutions of $\mathrm{^4He}$
in $\mathrm{^3He}$ is analyzed and discussed. Additional mechanism
called vacancy assisted nucleation is proposed. According to this
assumption the vacancy-impurity clusters should have b.c.c.
structure.
\end{abstract}


 \maketitle

\section{\label{sec:Intro} Introduction}

Vacancy-impurity clusters (VIC) were predicted in 1978 by this
author \cite{DP78} (see also Refs.~\onlinecite{Singapore, Nauka})
as one of a series structural effects that could appear due to the
quantum nature of vacancies and impurities in quantum crystals.
This effect was confirmed in 2001 \cite{Ganshin01} and further
investigated \cite{MaidanJLTP02,GanshinPhysB03,GanshinFNT03} in
the famous experiments of the Kharkov group, using new methods
based on a precise measuring of the pressure variations in phase
separation of solid solutions $\mathrm{^3He - ^4He}$. The effect
is based on the fact that the delocalization of a vacancy in a
quantum solid is attended with a lowering of the energy by an
amount of the half vacancion band width. This requires, however, a
good periodicity of the surrounding area. In helium solid
solutions the periodicity can be violated by isotope impurities.
Hence, the vacancy could i) push out the impurities, ii) attract
them and create a cluster of impurity atoms, or iii) rearrange
host atoms and impurities in a periodic lattice. We shaw that all
three possibilities can be realized in the helium solutions. Other
possibilities can be found in Refs.~\onlinecite{Singapore,Nauka}.
Ganshin et al.
\cite{Ganshin01,MaidanJLTP02,GanshinPhysB03,GanshinFNT03} observed
clusters of $\mathrm{^4He}$ impurities around vacancies in dilute
solutions of $\mathrm{ ^4He}$ in $\mathrm{^3He}$. They called them
Andreev--Pushkarov nanoclusters having probably in mind some
analogy between the structural effect in solid solutions and the
magnetic vacancy in pure $\mathrm{^3He}$ proposed by Andreev
\cite{Andreev76} but unfortunately not observed yet.

 The aim of this article is to consider some peculiarities of the
vacancy-impurity clusters and turn attention to an additional
mechanism in their creation.

\section{The rid effect}

Let us consider an isolated vacancy in a dilute solution. Let
$\varepsilon_0$ be the formation energy of a localized vacancy.
The delocalization lowers this energy by an amount of $\Delta/2$
where $\Delta$ is the vacancion energy band width. Since this is
only possible in a periodic lattice, the impurities should be
remouved from a region of radius $R$ (measured in this work in
interatomic distances $a = (V_m/N_A)^{1/3}$). Due to the
Heisenberg uncertainty principle the lowest energy level in such a
potential region is approximately
   \begin{equation}\label{eq:eps}
\varepsilon = \pi^2 \hbar^2/M R^2 = \pi^2 A/R^2
\end{equation}
where we used the relation between the vacancion effective mass
$M$ and the tunnelling amplitude $A$ for cubic crystals $ M =
\hbar^2/(A a^2)$ with the interatomic distance $a = 1$. In the
same notation the band width is $\Delta = z A$, $z$ being the
number of the nearest neighbours (for a h.c.p. lattice $ M =
\hbar^2/(2 A a^2)$, and the energy band consists of two partially
overlapping parts \cite{FNT75,Singapore,Nauka}).
 The ridding changes the entropy. Hence, the change of the
free energy is of the form
\begin{equation}\label{eq:free}
    F = E - TS = \varepsilon_0 -\frac{\Delta}{2} + \frac{\pi^2
    A}{R^2}+ \frac{4}{3}\pi R^3 T S
\end{equation}
where $S = x\ln (e/x)$ is the entropy per unit lattice site, and x
is the impurity concentration. Expression (\ref{eq:free}) has a
minimum at \cite{DP78}
 \begin{equation}\label{eq:R}
  R = \left(\frac{\pi A}{2 T S}\right)^{1/5}
\end{equation}
and the volume of the rid region is
\begin{equation}\label{eq:V_0}
    V_0 =\frac{4}{3}\pi \left(\frac{\pi A}{2 T S}\right)^{3/5}
\end{equation}
This result is restricted by the requirement that the free energy
be not smaller than $\varepsilon_0 - \Delta/2$, as well as by the
condition $R> a$. The latter condition is stronger, and reads $
TS< A$. In addition, the energy of the first quantum level
(\ref{eq:eps}) should be less than $\Delta/2$.

  Of course, expressions (\ref{eq:free}) - (\ref{eq:V_0}) are only
estimations showing the order of magnitude. These formulae do not
take into account the change of the impurity concentration in the
solution. The important result is that the size of the rid region
is determined by the interplay between the amplitude $A$ (the
vacancy exchange integral, not the band width) and the entropy
term and does not depend on the "won" energy due to the
delocalization, $\Delta/2$,  and the impuriton characteristics.
The quantum nature of the impuritons shows itself in their
mobility. The large value of the entropy term is the reason the
zone motion of defects in $\mathrm{^3He}$ to be suppressed.

 Analogous formula has been obtained by Andreev
\cite{Andreev76} for the number of aligned spins $N_s$ around a
vacancy in solid $\mathrm{^3He}$. In that case $S= n\ln 2 = \ln 2$
and the expression (\ref{eq:V_0}) takes the form
\begin{equation}\label{eq:Andre}
    N_s =\frac{4}{3}\pi \left(\frac{\pi A}{2 T \ln 2}\right)^{3/5}
\end{equation}
 This equation with $T= 150 \,
\mathrm{mK}$ and $A= 1 \,\mathrm{K}$ yields $R \approx 1.5$ and
$N_s \approx 14$. Note that with this value of $R$ the
delocalization energy $ \pi^2 A/R^2 \approx 4.5 \, \mathrm{K}$
turns out to be larger than the half band width $\Delta/2$ in
$\mathrm{^3He}$, i.e. the first quantum level does not lie in the
well. The minimum value of the free energy turns out to be higher
than the formation energy of a localaized vacancy: $F_{min} =
\varepsilon_0 -\Delta/2 + 11.85 \,A^{3/5} T^{2/5} \rightarrow
\varepsilon_0 -\Delta/2 + 5.55
> \varepsilon_0$. Therefore, the effect considered should take
place eventually in materials with very large bandwidths, but
hardly in $\mathrm{^3He}$.

Let us turn back to the solid solutions $\mathrm{^4He} -
\mathrm{^3He}$. The approximation above can be improved
\cite{Singapore,Nauka} for the more realistic case of vacancies in
a solutiowith a given small concentration $x_v= n_v/N$ and a
concentration of impurities $ x = n_i/N$. As a result of ridding,
the impurity concentration increases to the value
\begin{equation}\label{eq:x'}
    x' = \frac{n_i}{N - N x_v V} = \frac{x}{1- x_v V}
\end{equation}
where $V$ is the rid area around a vacancy. The problem can be
formulated as looking for the minimal work for changing the
concentration in the solution $ x \rightarrow x'$ by removing $x_v
V N$ particles of the solvent.

The Gibbs potential for a solution with concentration $x$ is of
the form \cite{Landau}
 \begin{equation}\label{eq:Gibbs1}
    \Phi_1 = N\mu_0 + N x T \ln \frac{x}{e} + Nx\psi
\end{equation}
where $\mu_0$ is the chemical potential of a pure solvent, and
$\psi=\psi(T,P)$.

 After ridding the impurity concentration becomes
$x'$, since the same number impurities is distributed over $N' = N
- x_v V N = N( 1 - x_v V) = N x'/x $ lattice sites. Therefore, the
Gibbs potential $\tilde\Phi_2 $ of the solution has the same form
(\ref{eq:Gibbs1}) with $x$ and $N$ replaced by $x'$ and $N'= N
x'/x$, while the potential of the pure solvent removed is $\Phi_0=
x_vVN\mu_0$. Their sum is
\begin{equation}\label{eq:Gibbstilde}
 \Phi_2 =  \tilde\Phi_2 +\Phi_0 = N\mu_0 + N x T \ln \frac{x'}{e} + Nx\psi
\end{equation}

The difference $\Phi_2 - \Phi_1$ is
\begin{equation}\label{DeltaFi}
\Delta \Phi = NTx\ln\frac{x'}{x}
\end{equation}
Hence, the variation of the entropy equals
\begin{equation}\label{DeltaS}
    \Delta S = Nx\ln\frac{x'}{x}
\end{equation}
 The change in the energy due to the presence of $Nx_v$ vacancions
in the system is
\begin{equation}\label{DelaE}
  \Delta E = N x_v  \left(\varepsilon_0 -\frac{\Delta}{2} + \frac{\pi^2
    A}{R^2}\right)
\end{equation}
and finally the free energy per particle takes the form
\begin{equation}\label{DeltaF}
    F = x_v  \left(\varepsilon_0 -\frac{\Delta}{2} + \frac{\pi^2
    A}{R^2}\right) - xT \ln\left( 1- x_v \frac{4}{3}\pi R^3\right)
\end{equation}
  The free energy $F$ has its minimum at \cite{Singapore,Nauka}
  \begin{equation}\label{eq:RR}
    R =\left(\frac{\pi A}{2 T x'}\right)^{1/5},
     \quad V_0 =\frac{4}{3}\pi\left(\frac{\pi A}{2 T x'}\right)^{3/5}
\end{equation}
 In (\ref{eq:RR}) the concentration $x'$ stands instead of the
 entropy $ S= x\ln(e/x)$. The minimal value of the free energy is
 then
 \begin{eqnarray}\label{Fmin}
F_{min} = \varepsilon_0 - \frac{\Delta}{2} +
\frac{10}{3}\pi\left(\frac{\pi
A}{2}\right)^{3/5}(Tx)^{2/5}\nonumber\\ = \varepsilon_0 -
\frac{\Delta}{2} + \frac{5}{2}n_x
\end{eqnarray}
where $n_x = x V$ is the number of impurities rid by one vacancy.
The change in the free energy affects the equilibrium vacancy
concentration:
\begin{eqnarray}\label{eq:v_eq}
    x_v &=& \exp\left\{ -\frac{\varepsilon_0 -\Delta/2} {T} -
    \frac{10\pi}{3} \left( \frac{\pi A}{2T}\right)^{3/5} x^{2/5}\right\}
    \nonumber\\
    &=& \bar x_v \exp\left\{-\frac{5}{2}V x\right\}.
\end{eqnarray}
Hence, the equilibrium vacancy concentration decreases with
increasing number of defects removed.

  Let us evaluate the volumes $V_0$ and $V$ for the typical values
$A= 1 \, \mathrm{K}$, $ T = 0.1 \, \mathrm{K}$, and $ x = 1$ \%.
Then one has by means of eqs. (\ref{eq:V_0}) and (\ref{eq:RR})
 $
    V_0 = 123 , \, \, V = 346.
 $
For $ T = 0.2 \, \mathrm{K}$ and $ x = 2$~\% the same expressions
yield $ V_0 = 60 \, $ and $ V = 150 $ (we did not take into
account an eventual dependence of $x'$ on $T$). The dependence of
the number of atoms in a cluster on temperature (at constant $x'$)
corresponding to formulae (\ref{eq:V_0}) and (\ref{eq:RR}) are
shown in Fig.~1.
 In fact, the number of
the removed impurities by a vacancy $n_x = x V \sim 2 \div 5$ and
the relative change of the concentration $\delta x/x $ is of the
order of several percents. However, this can lead to a noticeable
decreasing of the vacancy equilibrium concentration by more than 2
orders of magnitude (see eq. (\ref{eq:v_eq})).
\begin{figure} 
\includegraphics[width=2.7in]{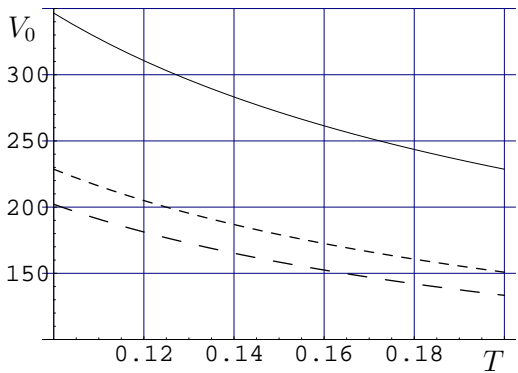}
\caption{\label{fig:Fig1} The number of atoms $V_0$ as versus
temperature at different concentrations. solid line: $x' = 1 \,
\%$ eq.~(\ref{eq:RR}), dashed line: $x' = 2 \, \%$
eq.~(\ref{eq:RR}), long-dashed line: $x' = 2 \, \%$
eq.~(\ref{eq:V_0}).}
\end{figure}
  It is worth noting that the region considered is not a cluster.
It consists of solvent atoms and has no boundaries, surface energy
etc.

\section{Diffusion}
 The ridding effect may be considered as an effective repulsion
which prevents impurities and vacancies be closer to each other.
This means that an effective cross-section of impuriton scattering
on vacancies can be introduced. It can be evaluated by means of
the radius $R$ and is of the order of $\sigma \sim \pi R^2 \sim
\pi (\pi A/2 T x)^{2/5}$.  In order to avoid any misunderstanding
we would like to note that this is not the case of $\mathrm{^4He}$
impurities in $\mathrm{^3He}$, because their zone motion is
embarrassed by the chaotic orientation of the nuclear spins of the
lattice atoms. It could take place in a strong orienting magnetic
field. The values of $\sigma$ for a hypothetic quasiparticle in a
medium with helium characteristics are $\sigma \sim 60 \, a^2$ (at
$T= 100$ mK and $x = 1\, \%$). The temperature and concentration
dependencies of the diffusion coefficient in this case are
\begin{equation}\label{D}
    D = D_0\frac{T^{2/5}}{x^{3/5}}
\end{equation}

  We considered till now concentrations less than the saturation
concentration $x_s$. If $x' > x_s$ then a new phase of impurities
appears. The effect of vacancies on this phase is considered in
the next section.

\section{Vacancy-impurity clusters}

 Let us consider a solid solution of impurity atoms (say $\mathrm{^4He}$)
with a concentration $x = n/N$ where $N$ is the number of solvent
atoms (say $\mathrm{^3He}$). If $ x < x_s$, Gibbs potential is of
the form (\ref{eq:Gibbs1})
\begin{equation}\label{eq:Fi1}
    \Phi_1 = N\mu_0 + n T \ln \frac{n}{Ne} + n\psi
\end{equation}
where $\mu_0$ refers to the pure solvent and $\psi= \psi(P,T)$.
The chemical potentials $\mu_3$ and $\mu_4$ that correspond to the
solvent atoms and impurities in the solution are
 \begin{equation}\label{mu34}
\mu_3 = \mu_0 - Tx, \quad \mu_4 = T \ln x + \psi
 \end{equation}

Let us move $\delta n$ impurities from the solution to their
"pure`` concentrated phase. The change of the thermodynamic
potential is
$$
 \Delta \Phi = -\delta n (T \ln x + \psi) + \delta n \mu_4^0
 $$
The chemical potential $\mu_4^0$ of the pure phase corresponds to
some saturation concentration, $ x_s$ (otherwise, the phase should
solve). Hence, $\mu_4^0 = T \ln x_s + \psi $, and
\begin{equation}\label{}
\Delta \Phi = -\delta n T \ln \frac{x}{x_s} = -NT\delta x  \ln
\frac{x}{x_s} .
\end{equation}
We suppose that all $\delta n$ atoms are confined to $n_v$
vacancies. Then, repeating the procedure in Sec.~II yields
\begin{equation}\label{eq:Rln}
    R = \left( \frac{\pi A}{2T|\ln(x/x_s)|}\right)^{1/5}
\end{equation}
with the restrictions $T|\ln(x/x_s)|< A$ and $x \ne x_s$. If the
concentration $x$ equals the saturation (equilibrium)
concentration then the entropy term vanishes and the vacancy is
fully delocalized. This effect should be seen in a rapid increase
of the vacancy mobility inside the pure phase. Such an observation
was reported in \cite{Goodkind} where fast nonphonon excitations
were found in solid $\mathrm{^4He}$ with a small concentration of
$\mathrm{^3He}$.
 Let us mention that
$x$ is the concentration in the matrix after the clusters are
formed. The values of $x$ and $x_s$ should be taken from the
experiment and the phase diagram. A quite sophisticated analysis
using known data for helium solutions was recently made in
Ref.~\cite{GanshinFNT03}. The authors found for the cluster size $
R \approx 2.1 \div 4.4 \, \mathrm{a}$. A small correction to these
values could appear if the band width $\Delta$ in the expression
for the localization energy $ \pi^2\Delta/R^2$ is replaced by $A$
(see eq. (\ref{eq:eps})). Otherwise, the first quantum level
$\epsilon$ turns out to lie out of the potential well with a depth
$\Delta/2$ \, ($\pi^2\Delta/R^2 > \Delta/2$ for all values of $R$
obtained).

For a rough evaluation one can simplify (\ref{eq:Rln}). If the
initial concentration is $x_0$, then $x= x_0 -x_v V - n_s/N
\approx x_s -x_v V $ where $n_s $ is the number of atoms in the
pure phase and we neglect the small number of solvent atoms solved
in the pure phase. The initial concentration drops out as it
should be in a case of phase separated mixture in equilibrium.
Then (\ref{eq:Rln}) can be rewritten in the form
\begin{equation}\label{R8}
  \frac{\pi A}{2T}=
  R^5 \ln |\left(1- \frac{x_v}{x_s}\frac{4}{3}\pi
  R^3\right)|\approx \frac{x_v}{x_s}\frac{4}{3}\pi
  R^8
\end{equation}
and therefore,
\begin{equation}\label{Rfin}
    R = \left(\frac{3A}{8T}\frac{x_s}{x_v}\right)^{1/8}
\end{equation}
For $ T= 150 \, \mathrm{mK}, \, x_v = 10^{-5}, \, x_s = 5\times
10^{-3}$ one has $ R \approx 2.5$ and $V \approx 60$ ($x_v V/x_s =
0.12 \ll 1$). Unfortunately this approximation requires $x_v V \ll
x_s $ and is not applicable to the most interesting low
temperature case where $x_s$ can be smaller than $x_v V$.

  It is worth noting, that VIC can appear both inside an
existing phase and as a separate inclusion. This makes difficult
to evaluate the role of the surface energy.

  \section{Local phase stratification effect, Vacancy assisted nucleation}

  This effect was predicted in the same work \cite{DP78} (see
Refs.~\cite{Singapore,Nauka} as well) and is another indication
for the quantum nature of vacancies. It appears when the vacancion
band widths in two phases of one and the same substance (e.g.
h.c.p. and b.c.c. $\mathrm{^4He}$) differ significantly. Let the
h.c.p. phase be the stable one (near the transition to b.c.c.). If
a vacancy appears, the energy increases by $ \epsilon_h -
\Delta_h/2$ (subscript $h$ corresponds to h.c.p.). If it were in a
b.c.c. phase, the energy change would be $ \epsilon_b -
\Delta_b/2$. Since near the phase separation the thermodynamic
potentials of the pure phases are almost equal, the more favorable
state is determined by the difference $\delta E= \epsilon_h -
\Delta_h/2 -(\epsilon_b - \Delta_b/2) = \delta \epsilon +
(\Delta_b- \Delta_h)/2$. For large enough $\Delta_b$ the
difference $\delta E >0$ independently of the sign of $\delta
\epsilon$ and the b.c.c. phase with a vacancy has lower energy. Of
course, the bottom of the band may become lower due to the large
difference of the vacancy formation energies, but this seems not
to be the case of helium crystals. The vacancy will rearrange the
$\mathrm{^4He}$ atoms from h.c.p. to b.c.c. hence creating a
nuclei of a new phase. The size $R$ of the nuclei can we evaluated
using the procedure applied above. The difference of the free
energy is
\begin{equation}\label{deltaF}
 \delta F =\delta \epsilon + \frac{\pi^2 A}{R^2} + T\delta S
 \frac{4}{3}\pi R^3
\end{equation}
where $A$ is the vacancion exchange integral in the b.c.c.
crystal. Having in mind that $T\delta S = q$ is the latent heat,
one obtains the nuclei radius
\begin{equation}\label{Rn}
    R_n = \left(\frac{\pi A}{2 q}\right)^{1/5}
\end{equation}
Let us note, that the surface energy in this case is smaller than
in the case when the matrix and the nuclei have different
structure (b.c.c. and h.c.p.). We were not able to find reliable
data for the latent heat at the temperature region considered. If
one takes $\Delta S \sim 10^{-2}$ extrapolating
\cite{EdwardsPandorf} then $q \sim 10^{-3}$ and $R_n \approx 4.4
$, $V_n \approx 350$.
  These are probably the clusters observed in the
experiments of Kharkov group. After the inclusion of
$\mathrm{^4He}$ with a vacancy appears, a local phase transition
to the more favorable b.c.c. phase is performed. This makes the
VIC even more stable. We suppose, that this effect could
contribute to the curious observation in \cite{Ehrlich} that the
lattice constants in both phases of a phase separated solution
increase after heating.

Experimental evidences to support our hypothesis should be looked
for not only in a phase separated solutions. As we showed in
\cite{DP78,Singapore,Nauka} inclusions of b.c.c. $\mathrm{^4He}$
should manifest themselves in a hysteresis at b.c.c.-h.c.p.
transitions of pure $\mathrm{^4He}$. In this case one can use the
experimental data for the transition line h.c.p.-b.c.c. and a
nuclei contains $\sim 200\div 250$ atoms. Phase nuclei of the same
kind and of approximately the same size should appear in the
liquid helium near the liquid-solid transition line.

  The theory of VIC is instructive from another point of view as
well. At the first glance, the small exchange integral in
spin-spin interaction of $\mathrm{^3He}$ atoms compared to the
vacancion bandwidth should lead to their easy orientation. It does
not happen due to the large entropy factor. So, vacancy prefers to
rearrange atoms (not spins) organizing a local phase of impurities
and to delocalize in this phase. Its choice is the minimal change
of entropy.

\section{Conclusion}
 In this work three effects due to the quantum nature of
vacancies in solid helium were considered relative to the
vacancy-impurity complexes observed recently in phase separated
solid solutions of $\mathrm{^4He}$ in $\mathrm{^3He}$. The size of
a cluster depends on the ratio of the vacancy exchange integral
$A$ and the entropy term $T\Delta S$, and does not depend directly
on the vacancion band width. The experimental results cannot give,
therefore, information about the band width. Nevertheless, it is
essential for the phase transition. It is supposed that VIC in
rear $\mathrm{^4He}$~-~$\mathrm{^3He}$ solutions undergo two
transitions: first complexes with h.c.p. structure appear and then
they transfer to b.c.c. clusters. The transition is controlled by
the positions of the bottoms of the vacancy bands, and by the
minimum entropy change. The evaluation of the clusters size is in
good agreement with the experimental observations. It is shown,
that near the separation line the size of the vacancy cluster
rapidly increases, the vacancy becomes delocalized and mobil
inside the $\mathrm{^4He}$- phase. It is supposed that this could
be the fast nonphonon mode observed in \cite{Goodkind}. The effect
of vacancy assisted local phase stratification should be
observable near the b.c.c.-h.c.p. as well as b.c.c.-liquid
transition.


\begin{acknowledgments}
I am much obliged to Prof. R. Dandoloff for extended discussions
and collaboration in preparing this work. I am thankful to CNRS
and Prof.~T.H.~Diep for the support and hospitality.
Acknowledgments are due to the partial financial support from the
National Science Fund, Contract F-911.
\end{acknowledgments}

\end{document}